\documentclass[amsmath,amssymb,showpacs,twocolumn]{revtex4-1}
\usepackage{psfrag,tabls,bm,bbm,color,graphicx,amsopn,dsfont}

\usepackage{epstopdf}
\begin{document}

\title{Extension of the Nakajima-Zwanzig approach to multitime correlation functions 
of open systems}

\author{Anton Ivanov}
\email[]{anton.ivanov@physik.uni-freiburg.de}
\affiliation{Physikalisches Institut, Universit\"at Freiburg, Herrmann-Herder-Stra{\ss}e 3, D-79104 Freiburg, Germany}
\author{Heinz-Peter Breuer}
\affiliation{Physikalisches Institut, Universit\"at Freiburg, Herrmann-Herder-Stra{\ss}e 3, D-79104 Freiburg, Germany}

\date{\today}

\begin{abstract}

We extend the Nakajima-Zwanzig projection operator technique to the determination of 
multitime correlation functions of open quantum systems. The correlation functions are 
expressed in terms of certain multitime homogeneous and inhomogeneous memory kernels 
for which suitable equations of motion are derived. We show that under the condition of finite
memory times these equations can be used to determine the memory kernels by employing
an exact stochastic unraveling of the full system-environment dynamics. The approach
thus allows to combine exact stochastic methods, feasible for short times, with long-time
master equation simulations. The applicability of the method is demonstrated by 
numerical simulations of 2D-spectra for a donor-acceptor model, and by comparison
of the results with those obtained from the reduced hierarchy equations of motion.
We further show that the formalism is also applicable to the time evolution of a periodically 
driven two-level system initially in equilibrium with its environment.

\end{abstract}


\maketitle

\section{Introduction}
\label{sec:intro}

The simulation of the dynamics of an open quantum system coupled to an infinitely large environment \cite{TheWork} is still a problem that attracts a lot of attention since there is a need for the development of reliable and fast numerical methods.
Approaches based on Redfield- or Lindblad-like equations (see, e.g., \cite{Vulto_1999, Jang_2002, Jang_2004, Mohseni_2008, Castro_2008, Braig_2003, Mitra_2004, Koch_2005, Haenggi_2013}),
(self-consistent) perturbation expansions in some small parameter 
within the Keldysh formulation \cite{Aligia_2006, Monreal_2010} or projection operator techniques
\cite{Nakajima,Zwanzig,Shibata}
do not cover the whole system parameter range of interest.
These gaps can be filled by the use of numerically expensive approaches like time dependent density matrix numerical renormalization groups \cite{Daley_2004, White_2004, Schmitteckert_2004}, multilayer multi-configuration time-dependent Hartree method in second quantisation representation \cite{Wang_2003,Wang_2011}, real time quantum Monte Carlo methods \cite{Muehlbacher_2008, Muehlbacher_2004} and iterative path summation schemes \cite{Weiss_2008,Segal_2010}.

Although being exact, their computational cost increases exponentially in time, which often requires their combination with other methods in order to obtain the stationary state of the system. In \cite{Wilner_2013,Cohen_2011} the Nakajima-Zwanzig generalised quantum master equation is used to extract the specific memory kernels from the early time evolution of the system, which was initially obtained by the use of one of the exact approaches mentioned above. The memory kernels are then used for the calculation of the system dynamics for arbitrary long times. 


The advances in nonlinear optical spectroscopy increased the need to develop efficient methods for calculating system multitime correlation functions. 

One of the important examples is the two-dimensional (2D) spectroscopy of a photosynthetic
pigment-protein complex known as Fenna-Mathew-Olsen
(FMO) complex, which is obtained from the knowledge of two three-time correlation functions \cite{Engel_2007}. 


In this work we extend the Nakajima-Zwanzig projection operator approach to the calculation of multitime correlation functions (MTCF), which requires the introduction of multitime homogeneous and inhomogeneous kernels. By having the information about the kernels in some finite time range we are able to calculate the MTCFs for an arbitrary set of times. We will see that the formalism can also be applied to problems which, at first glance, do not require the calculation of MTCFs, namely the time evolution of a periodically driven system being initially in equilibrium with its environment. 

In order to calculate the required multitime kernels we construct a set of equations. If the problem can be solved efficiently by the hierarchy equations of motion (HEOM) method \cite{Tanimura_06,Tanimura_09}, then the input information for the equations can be obtained by slight modification of the method. Otherwise one can use a stochastic unravelling approach which is well suited for this task as long as the kernels decay to zero sufficiently fast.

The main advantage of this two-step approach is that it gives us the possibility to calculate MTCFs for problems that can not be described by HEOM. Even if the HEOM method is applicable it can be still more efficient to calculate the multitime kernels via the HEOM method and then the multitime propagators, than the direct calculation of the MTCFs.

The paper is structured as follows. 
In Sec.~\ref{subsec:N-Zwanzig_derivation} we derive briefly the solution of the Nakajima-Zwanzig equation. The generalisation of the problem to multitime correlation functions is presented in Sec.~\ref{sec:M-time_corr_fct}, and in Sec.~\ref{sec:Eq_for_multitime_kernels} we derive the rules for constructing equations for the multitime kernels. 
In Sec.~\ref{subsec:Per_driv_sys} we apply the formalism to a periodically driven system being initially in equilibrium with its environment. The input information needed for the solution of the equations for the multitime kernels is calculated by use of a stochastic unravelling approach, which is presented in Sec.~\ref{subsec:Stoch_unravel}. The reliability of the method is tested in Sec.~\ref{sec:Results}. In  Sec.~ \ref{subsec:Results_per_driven_sys} the numerical results for the problem proposed in Sec.~ \ref{subsec:Per_driv_sys} are presented, and in 
Sec.~\ref{subsec:2D_spec_fct} the 2D-spectra of a donor-acceptor model is calculated. In both cases the results are compared with those obtained from the HEOM approach. Finally, conclusions about the advantages and drawbacks of the method are given in Sec.~\ref{sec:Conclusion}.

\section{Theory} \label{sec:II}
\subsection{Nakajima-Zwanzig equation}
\label{subsec:N-Zwanzig_derivation}
We consider an open system $S$ coupled to some bath $B$. The total Hilbert space is $\mathcal{H} = \mathcal{H}^{}_S{} \otimes \mathcal{H}^{}_{B} $. The Liouvillian operators, that describe the system, the bath and the system-bath coupling, are denoted by $\mathcal{L}^{}_{S}$, $\mathcal{L}^{}_{B}$ and $\mathcal{L}^{}_{SB}$, respectively. 
The Liouvillian for the total system is thus given by
\begin{equation}
 \mathcal{L} =
 \mathcal{L}^{}_{S} + \mathcal{L}^{}_{B} + \mathcal{L}^{}_{SB}.
\end{equation}
For simplicity we assume that all operators do not depend on time, but the results can be extended for time dependent Liouvillians as well. We also define the projection operators 
$\mathcal{P} = {\rm tr^{}_{B} } [ \ldots ] \otimes R$ and $\mathcal{Q} = 1 - \mathcal{P}$, where ${\rm tr^{}_{B} } [ \ldots ]$ denotes the trace over the bath degrees of freedom and 
$R$ is an arbitrary density operator for the bath with the properties ${\rm tr^{}_{B}}[R] = 1$ and $\mathcal{L}^{}_{B} R = 0$.

In the following we will often use the identity
\begin{subequations}
\allowdisplaybreaks
\begin{align}
\hat{T} \Big[ 
		e^{\int^{t}_{ t^{}_{0} } ds ( B  (s) + C (s) ) }_{} \Big] =  \hspace{50.0mm} \nonumber\\ =
		\hat{T} \Big[   
		e^{\int^{t}_{ t^{}_{0} }  ds B (s) }_{}		
		+ \int^{t}_{ t^{}_{0} } d\tau e^{ \int^{t}_{\tau^{}_{}} ds B (s) }_{} C (\tau) 
		e^{ \int^{\tau}_{t^{}_{0}} d\tilde{s} ( B (\tilde{s}) + C (\tilde{s}) ) }_{}
		\Big] \label{eq:SD_Identity_A} \\
 =
		\hat{T} \Big[   
		e^{ \int^{t}_{ t^{}_{0} }  ds B (s) }_{}		
		+ \int^{t}_{ t^{}_{0} } d\tau 
		e^{ \int^{ t }_{\tau^{}_{}} ds ( B (s) + C (s) ) }_{}
		C (\tau) e^{ \int^{\tau}_{t^{}_{0}} d\tilde{s} B (\tilde{s}) }_{}
		\Big], \label{eq:SD_Identity_B}
\end{align}
\end{subequations}
where $B(t)$ and $C(t)$ are any time dependent superoperators and
$\hat{T}$ is the time ordering operator. The pair $(B,C)$ will be replaced by $( \mathcal{P}\mathcal{L}, \mathcal{Q} \mathcal{L})$, $(\mathcal{L}\mathcal{P}, \mathcal{L} \mathcal{Q} )$ or $( \mathcal{Q} \mathcal{L}, \mathcal{P} \mathcal{L})$, $( \mathcal{L} \mathcal{Q}, \mathcal{L} \mathcal{P} )$. The last set of identities that we will need is
\begin{equation}
\begin{array}{ll}
Be^{\mathcal{L}Bt}_{} = e^{B \mathcal{L}t}_{} B,  	&  B\in \lbrace \mathcal{P},\mathcal{Q} \rbrace, \\
Be^{C\mathcal{L}t}_{} = e^{\mathcal{L} Ct}_{} B = B, &  (B,C) \in \lbrace (\mathcal{P},\mathcal{Q}),(\mathcal{Q},\mathcal{P}) \rbrace .
\end{array}
\end{equation}
The density matrix of the open system $ \rho^{}_{S} (t) $ is given by
\begin{align}
\rho^{}_{S}  (t)  & = {\rm tr^{}_{B}} \big[  e^{\mathcal{L} t } \rho^{}_{0} \big] = U (t) \rho^{}_{S} (0) + V (t) , \label{eq:Def_rho_s_B}
\end{align}
where the initial state of the total system is denoted by $\rho^{}_{0}$ and  $U$, $V$ represent 
the homogeneous and the inhomogeneous propagators, respectively:
\begin{subequations}
\begin{align}
U (t) & = {\rm tr^{}_{B}} \big[ e^{\mathcal{L} t }_{} R \big], \label{eq:Def_U} \\
V (t) & = {\rm tr^{}_{B}} \big[ e^{\mathcal{L} t }_{} \mathcal{Q} \rho^{}_{0} \big]. \label{eq:Def_V}
\end{align}
\end{subequations}

By applying Eq.~\eqref{eq:SD_Identity_A} for $(B,C) = (\mathcal{L}\mathcal{P} ,\mathcal{L}\mathcal{Q} )$ to $U (t) = {\rm tr^{}_B} \big[ e^{\mathcal{L}t}_{} \mathcal{P} R \big]$, substituting $\mathcal{P} R X = RX$ for every operator $X$ acting on $\mathcal{H}^{}_S$ and then using Eq.~\eqref{eq:Reduction_rules_B} we obtain
\begin{equation}
\label{eq:Nak_Zwanz_Upart_B}
\begin{array}{l}
U (t) =  {\rm tr^{}_{B} } \big[  e^{ \mathcal{L} \mathcal{P} t }_{} R  \big] \\
 + \int^{t}_{0} d\tau \int^{\tau}_{0} d\tau'   
	{\rm tr^{}_{B} } \big[ e^{ \mathcal{L} \mathcal{P} (t-\tau)  }_{} \mathcal{L} \mathcal{Q} e^{ \mathcal{L} \mathcal{Q} (\tau - \tau') }_{} \mathcal{L} \mathcal{P} e^{\mathcal{L} \tau'}  R \big].
\end{array}
\end{equation}
Since for every operator $X$ acting on $\mathcal{H}$ we have
\begin{equation}
\begin{array}{rcl}
{\rm tr^{}_{B}} [ \mathcal{L} \mathcal{P} X ] & = &  \mathcal{L}^{}_{\bar{S}}  {\rm tr^{}_{B} } [X],
\end{array}
\end{equation}
where $\mathcal{L}^{}_{\bar{S}} \equiv \mathcal{L}^{}_{S} + \langle \mathcal{L}^{}_{SB} \rangle $ with $\langle \mathcal{L}^{}_{SB} \rangle \equiv {\rm tr^{}_{B}}[\mathcal{L}^{}_{SB} R ]$, it follows that
\begin{subequations}
\begin{align}
 {\rm tr^{}_{B}} \big[ e^{ \mathcal{L} \mathcal{P} t }_{} X \big] & = 
 u^{}_{ \bar{S} } (t)  {\rm tr^{}_{B}} \big[ X \big],\\
u^{}_{\bar{S}} (t) & = e^{  \mathcal{L}^{}_{ \bar{S}} t }_{} .
\end{align}
\end{subequations}
Equation \eqref{eq:Nak_Zwanz_Upart_B} can then be rewritten as
\begin{equation}
 U (t) = u^{}_{\bar{S} } (t) +
\int^{t}_{0} d\tau \int^{ \tau }_{0} d\tau' u^{}_{ \bar{S} }  (t-\tau) K  (\tau - \tau')  U  (\tau') ,
\label{eq:Def_U_wieder}
\end{equation}
where the memory kernel is given by
\begin{equation}
 K (t) = { \rm tr^{}_{B} } \big[ \mathcal{L}  e^{ Q\mathcal{L}t }_{} Q\mathcal{L}R \big]   
  = { \rm tr^{}_{B} } \big[ \mathcal{L}^{}_{SB}  e^{ Q\mathcal{L}t }_{} 
  Q\mathcal{L}^{}_{SB} R \big]. \label{eq:Def_K__with_Lsb}
\end{equation}
The last expression of the previous equation is obtained by use of 
$\mathcal{L}^{}_{B} R = 0$ and of the relation 
${\rm tr^{}_{B}}[ \mathcal{Q} X] = 0$ which holds for all
operators $X$ acting on $\mathcal{H}$.

By the use of Eq.~\eqref{eq:SD_Identity_B} for $(B,C) = ( \mathcal{Q} \mathcal{L}, \mathcal{P} \mathcal{L} )$ and ${\rm tr^{}_{B}} [ e^{Q\mathcal{L} t }_{} \mathcal{Q} X ] = 0 $ 
the equation for the inhomogeneous propagator [Eq.~\eqref{eq:Def_V}] becomes
\begin{subequations}
\begin{align}
V (t) &=  \int^{t}_{0} d\tau U (t - \tau)  I  (\tau ) ,  \label{eq:Def_V_wieder} \\
I (t) &= { \rm tr^{}_{B} } \big[ \mathcal{L}          e^{ Q\mathcal{L}t }_{} Q\rho^{}_{0} \big]  
 = { \rm tr^{}_{B} } \big[ \mathcal{L}^{}_{SB}  e^{ Q\mathcal{L}t }_{} Q\rho^{}_{0} \big]. \label{eq:Def_I__with_Lsb}
\end{align}
\end{subequations}
The kernel $I (t) $ is also known as inhomogeneity.
Combining Eqs.~\eqref{eq:Def_rho_s_B} and \eqref{eq:Def_V_wieder} and then using 
Eq.~\eqref{eq:Def_U_wieder} we obtain the solution of the Nakajima-Zwanzig equation:
\begin{eqnarray} \label{eq:N_Zwanzig_solution}
\rho^{}_{S} (t) &=& u^{}_{\bar{S}} (t) \rho^{}_{S}  (0) +
		\int^{t}_{0} d\tau u^{}_{ \bar{S} } (t-\tau) I  ( \tau ) \\
		&& + \int^{t}_{0} d\tau \int^{ \tau }_{0} d\tau' u^{}_{\bar{S}} (t-\tau) K  (\tau - \tau') \rho^{}_{S}  (\tau'). \nonumber
\end{eqnarray}


At first glance it seems unnecessary to express $\rho^{}_{S} {\scriptstyle  }$ in terms of $U$ and $V$ as it is done in Eq. \eqref{eq:Def_rho_s_B}, because for the time evolution of the system density matrix it is sufficient to solve only Eq. \eqref{eq:N_Zwanzig_solution}. The alternative form in Eq. \eqref{eq:Def_rho_s_B} is preferred if we are interested in the calculation of multitime correlation functions.

\subsection{Multitime correlation functions }
\label{sec:M-time_corr_fct}

The multitime correlation function of an arbitrary set of system operators $ \lbrace A^{}_{j} \rbrace^{}_{j \in \mathbb{N}}$ applied at times 
$t^{}_{j \ldots 1} \equiv t^{}_{j} + \ldots + t^{}_{1} $ is given by
\begin{eqnarray}
 \lefteqn{
 \langle A^{}_{N} {\scriptstyle  (t^{}_{N\ldots 1}) } \ldots A^{}_{1} {\scriptstyle ( t^{}_{1} ) } 
 \rangle } \nonumber \\
 &=& {\rm tr^{}_{S}} \big[ 
		A^{}_{N}  U^{A^{}_{N-1} \ldots A^{}_{1}}_{ (t^{}_{N}, \ldots, t^{}_{1} ) }  
		\rho^{}_{S} (0)  			
		+ A^{}_{N} V^{A^{}_{N-1} \ldots A^{}_{1}}_{(t^{}_{N}, \ldots, t^{}_{1} )}
		\big],
\end{eqnarray}
where we have introduced the $N$-time homogeneous and inhomogeneous propagators:
\begin{subequations}
\begin{align}
 U^{A^{}_{N-1}  \ldots A^{}_{1}}_{ (t^{}_{N}, \ldots, t^{}_{1})} &= {\rm tr^{}_{B}} \big[ e^{\mathcal{L} t^{}_{N} }_{}  A^{}_{N-1}  \ldots e^{\mathcal{L} t^{}_{2} }_{}  A^{}_{1}   e^{\mathcal{L} t^{}_{1} }_{}  R \big], \\
V^{A^{}_{N-1}  \ldots A^{}_{1}}_{ (t^{}_{N}, \ldots, t^{}_{1})} &= {\rm tr^{}_{B}} \big[ e^{\mathcal{L} t^{}_{N} }_{}  A^{}_{N-1}  \ldots e^{\mathcal{L} t^{}_{2} }_{}  A^{}_{1}   e^{\mathcal{L} t^{}_{1} }_{}  \mathcal{Q} \rho^{}_{0} \big].
\end{align}
\end{subequations}
Every $N$-time propagator can be expressed as a function of $n$-time kernels 
with $n \leq N$ and of the propagators $U$ and $V$. The multitime homogeneous $K^{A^{}_{N-1} \ldots A^{}_{1} }_{}$ and inhomogeneous $I^{A^{}_{N-1} \ldots A^{}_{1} }_{}$ kernels are defined as
\begin{subequations}
\label{eq:Def_K_and_I_multitime}
\begin{flalign}
 & K^{ A^{}_{N-1} \ldots A^{}_{1} }_{ ( t^{}_{N}, \ldots , t^{}_{1} )} =  \label{eq:Def_K_multitime}\\
 & { \rm tr^{}_{B} } \big[ \mathcal{L}^{}_{SB} e^{ Q\mathcal{L}t^{}_{N} }_{} Q A^{}_{N-1} \ldots e^{ Q\mathcal{L}t^{}_{2} }_{} Q A^{}_{1}  e^{ Q\mathcal{L}t^{}_{1} }_{} Q\mathcal{L}^{}_{SB} R \big], \nonumber\\
 & I^{ A^{}_{N-1} \ldots A^{}_{1} }_{ ( t^{}_{N}, \ldots , t^{}_{1} )} =  \label{eq:Def_I_multitime}\\
 & { \rm tr^{}_{B} } \big[ \mathcal{L}^{}_{SB} e^{ Q\mathcal{L}t^{}_{N} }_{} Q A^{}_{N-1} \ldots e^{ Q\mathcal{L}t^{}_{2} }_{} Q A^{}_{1}  e^{ Q\mathcal{L}t^{}_{1} }_{} Q \rho^{}_{0} \big]. \nonumber
\end{flalign}
\end{subequations}
The procedure to obtain the desired expressions can be entirely summarised in applying the following reduction rules:
\begin{subequations}
\allowdisplaybreaks
\label{eq:Reduction_rules}
\begin{align}
\mathcal{B} e^{ \mathcal{L} t }_{} \mathcal{Q} & =  \int^{t}_{0} d\tau \mathcal{B}  e^{ \mathcal{L} (t-\tau) }_{} \mathcal{P}\mathcal{L} e^{ \mathcal{Q} \mathcal{L} \tau }_{} \mathcal{Q}, \label{eq:Reduction_rules_A}\\
\mathcal{Q} e^{ \mathcal{L} t }_{} \mathcal{P} & =  \int^{t}_{0} d\tau \mathcal{Q} e^{ \mathcal{L} \mathcal{Q} (t-\tau) }_{}  \mathcal{L}\mathcal{P}  e^{ \mathcal{L} \tau }_{} \mathcal{P}, \label{eq:Reduction_rules_B}\\
\mathcal{Q} e^{ \mathcal{L} t }_{} \mathcal{Q} & =  
\mathcal{Q}  e^{ \mathcal{L} \mathcal{Q} t }_{} \mathcal{Q} \label{eq:Reduction_rules_C}\\
 & 
+ \int^{t}_{0} d\tau \int^{\tau}_{0} d\tilde{\tau}   \mathcal{Q} e^{ \mathcal{L} \mathcal{Q} (t-\tau) }_{} \mathcal{L} \mathcal{P}  e^{  \mathcal{L} (\tau - \tilde{\tau}) }_{} \mathcal{P} \mathcal{L} e^{ \mathcal{Q} \mathcal{L} \tilde{\tau} }_{} \mathcal{Q}, \nonumber\\
\mathcal{Q} e^{ \mathcal{L} t }_{} \mathcal{Q} \rho^{}_{0} & = 
\mathcal{Q} e^{ \mathcal{L} \mathcal{Q} t }_{} \mathcal{Q} \rho^{}_{0}  \label{eq:Reduction_rules_D}\\
& + \int^{t}_{0} d\tau   \mathcal{Q} e^{ \mathcal{L} \mathcal{Q} (t-\tau) }_{} \mathcal{L} \mathcal{P} e^{ \mathcal{L} \tau }_{}  \mathcal{Q} \rho^{}_{0},  \nonumber
\end{align}
\end{subequations}
where $ \mathcal{B} \in \lbrace  \mathcal{P}, {\rm tr^{}_{B} } \rbrace $. In addition, one always has to decompose the system operators as $A^{}_{} = \mathcal{P} A^{}_{} \mathcal{P} + \mathcal{Q} A^{}_{} \mathcal{Q} $.

We will apply these rules for the two- and three-time propagators. Starting from
\begin{subequations}
\label{eq:gen_red_Expressioin}
\begin{align}
{\rm tr^{}_{B}} \big[ e^{ \mathcal{L} t^{}_{2} }_{} A^{}_{1} e^{ \mathcal{L} t^{}_{1} }_{} \mathcal{B} \big], & \hspace{4.0mm}  \mathcal{B} \in \lbrace \mathcal{P} R ,\mathcal{Q} \rho^{}_{0} \rbrace, \\
{\rm tr^{}_{B}} \big[ e^{ \mathcal{L} t^{}_{3} }_{} A^{}_{2} e^{ \mathcal{L} t^{}_{2} }_{} A^{}_{1} e^{ \mathcal{L} t^{}_{1} }_{} \mathcal{B} \big], & \hspace{4.0mm} \mathcal{B} \in \lbrace \mathcal{P} R, \mathcal{Q}\rho^{}_{0} \rbrace,
\end{align}
\end{subequations}
and applying the reduction rules for $\mathcal{B}=\mathcal{P} R$ we obtain the following equations:
\begin{subequations}
\allowdisplaybreaks
\label{eq:Homog_prop_multitime}
\begin{align}
U^{A^{}_{1} }_{(t^{}_{2}, t^{}_{1})} & = U {\scriptstyle (t^{}_{2}) } A^{}_{1} U {\scriptstyle (t^{}_{1}) } \label{eq:Homogen_prop_multitime_first}\\
	& + U {\scriptstyle (t^{}_{2} - \tau^{}_{2} ) } K^{A^{}_{1}}_{(\tau^{}_{2}, \tau^{}_{1})} U {\scriptstyle (t^{}_{1} - \tau^{}_{1}) } , \nonumber \\
U^{A^{}_{2} A^{}_{1} }_{(t^{}_{3},t^{}_{2}, t^{}_{1})} & = 
	U {\scriptstyle (t^{}_{3}) } A^{}_{2} U {\scriptstyle (t^{}_{2}) } A^{}_{1} U {\scriptstyle (t^{}_{1})} \nonumber\\
	& + U {\scriptstyle (t^{}_{3})} A^{}_{2} U { \scriptstyle (t^{}_{2}-\tau^{}_{2})} K^{ A^{}_1 }_{( \tau^{}_2, \tau^{}_1 )} U { \scriptstyle (t^{}_{1} - \tau^{}_1 )}  \nonumber\\
	&  +  U { \scriptstyle (t^{}_{3} - \tau^{}_3 )} K^{ A^{}_2 }_{( \tau^{}_3 , \tau^{}_2 )} U {\scriptstyle( t^{}_{2} - \tau^{}_{2} )} A^{}_1  U { \scriptstyle ( t^{}_{1} )} \nonumber\\ 
	&  +  U { \scriptstyle (t^{}_{3} - \tau^{}_3 ) } K^{ A^{}_2 }_{( \tau^{}_3 , \tau^{}_2 )} U { \scriptstyle (t^{}_{2} - \tau^{}_{2} - \tau'^{}_{2})} K^{ A^{}_1 }_{( \tau'^{}_2, \tau^{}_1 )} U { \scriptstyle (t^{}_{1} - \tau^{}_1 )} \nonumber \\
	&  +  U { \scriptstyle (t^{}_{3} - \tau^{}_3 )} K^{ A^{}_2 A^{}_1 }_{( \tau^{}_3 , t^{}_2, \tau^{}_1 )} U { \scriptstyle ( t^{}_{1} - \tau^{}_{1} )} \nonumber\\
	&  =  U { \scriptstyle (t^{}_{3})} A^{}_{2} U^{A^{}_{1}}_{(t^{}_{2}, t^{}_{1})}   \label{eq:Homogen_prop_multitime_second} \\
	&  +  U { \scriptstyle (t^{}_{3} - \tau^{}_{3} )} K^{A^{}_{2}}_{(\tau^{}_{3}, \tau^{}_{2})} U^{A^{}_{1}}_{(t^{}_{2} - \tau^{}_{2}, t^{}_{1})} \nonumber \\
	&  +  U { \scriptstyle (t^{}_{3} - \tau^{}_{3} )} K^{A^{}_{2}A^{}_{1}}_{(\tau^{}_{3}, t^{}_{2}, \tau^{}_{1})} U { \scriptstyle (t^{}_{1} - \tau^{}_{1})}. \nonumber
\end{align}
\end{subequations}
Here and in all other equations we integrate over all $\tau^{}_{j}$ variables from $0$ to $t^{}_j$ and over all $\tau'^{}_j$ variables from $0$ to $t^{}_j - \tau^{}_j$ if the integration range is not shown explicitly.

\begin{figure}
	\centering
	\includegraphics[width=.95\linewidth]{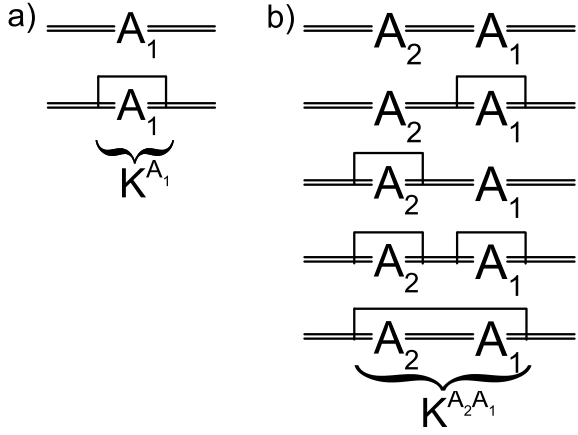}
	\caption{Diagramatic representation of the terms contributing to Eq. \eqref{eq:Homogen_prop_multitime_first} (a) and to Eq. \eqref{eq:Homogen_prop_multitime_second} (b). The thick double line refers to the 
	U-propagator.}
	\label{fig:Homog_diag}
\end{figure}

\begin{figure}
	\centering
	\includegraphics[width=.95\linewidth]{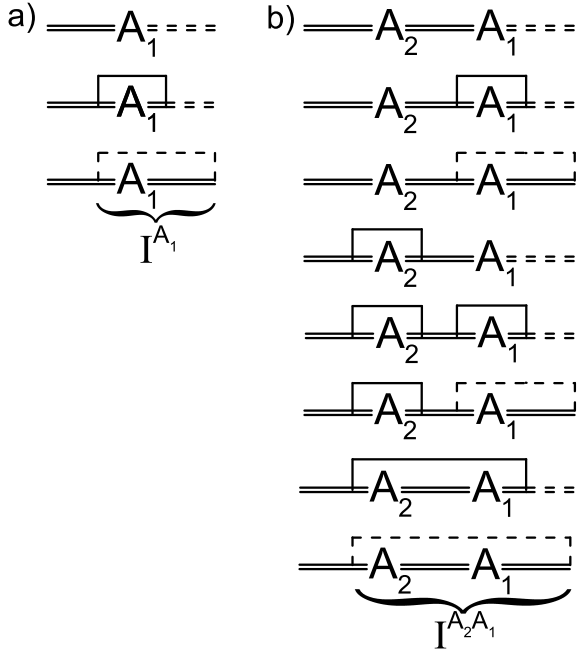}
	\caption{Terms contributing to Eq. \eqref{eq:Inhomogen_prop_multitime_first} (a) and to Eq. \eqref{eq:Inhomogen_prop_multitime_second} (b). The dashed double line refers to the $V$-propagator.}
  \label{fig:Inhomog_diag}
\end{figure}

The desired equations for the two- and three-time inhomogeneous propagators are obtained by applying the reduction rules to Eq.  \eqref{eq:gen_red_Expressioin} for $\mathcal{B} = \mathcal{Q}\rho^{}_0$:
\begin{subequations}
\allowdisplaybreaks
\label{eq:Inhomogen_prop_multitime}
\begin{align}
V^{A^{}_{1} }_{(t^{}_{2}, t^{}_{1})} & =  U { \scriptstyle (t^{}_{2})} A^{}_{1} V { \scriptstyle (t^{}_{1})} \label{eq:Inhomogen_prop_multitime_first}\\
	& + U { \scriptstyle (t^{}_{2} - \tau^{}_{2} )} K^{A^{}_{1}}_{(\tau^{}_{2}, \tau^{}_{1})} V { \scriptstyle (t^{}_{1} - \tau^{}_{1})} \nonumber \\
	& + U { \scriptstyle (t^{}_{2} - \tau^{}_{2} )} I^{A^{}_{1}}_{ (\tau^{}_{2}, t^{}_{1} ) } , \nonumber\\
V^{A^{}_{2} A^{}_{1} }_{( t^{}_{3}, t^{}_{2}, t^{}_{1})} 
	& = U {\scriptstyle (t^{}_{3})} A^{}_{2} U { \scriptstyle (t^{}_{2})} A^{}_{1}  V { \scriptstyle ( t^{}_{1} ) } \nonumber \\
	& + U {\scriptstyle (t^{}_{3})} A^{}_{2} U { \scriptstyle (t^{}_{2} - \tau^{}_2)} K^{ A^{}_1 }_{ ( \tau^{}_2, \tau^{}_{1} ) } V { \scriptstyle (t^{}_{1} - \tau^{}_1 ) }  \nonumber\\
	& + U { \scriptstyle (t^{}_{3})} A^{}_{2} U { \scriptstyle (t^{}_{2} - \tau^{}_2)} I^{ A^{}_1 }_{ ( \tau^{}_2, t^{}_{1}    ) }  \nonumber\\
	& + U { \scriptstyle (t^{}_{3} - \tau^{}_3 )} K^{ A^{}_2 }_{ (\tau^{}_3, \tau^{}_2 ) } U {\scriptstyle (t^{}_{2} - \tau^{}_2 )} A^{}_{1}  V { \scriptstyle ( t^{}_{1} ) }  \nonumber\\
	& + U { \scriptstyle (t^{}_{3} - \tau^{}_3 )} K^{ A^{}_2 }_{ (\tau^{}_3, \tau^{}_2 ) } U { \scriptstyle (t^{}_{2} - \tau^{}_2 - \tau'^{}_2 )} K^{ A^{}_1 }_{ ( \tau'^{}_2, \tau^{}_{1} ) } V { \scriptstyle (t^{}_{1} - \tau^{}_1 ) }  \nonumber\\
	& + U { \scriptstyle (t^{}_{3} - \tau^{}_3 )} K^{ A^{}_2 }_{ (\tau^{}_3, \tau^{}_2 ) } U {\scriptstyle (t^{}_{2} - \tau^{}_2 - \tau'^{}_2 )} I^{ A^{}_1 }_{ ( \tau'^{}_2, t^{}_{1}    ) }  \nonumber\\
	& + U { \scriptstyle (t^{}_{3} - \tau^{}_3 )} K^{ A^{}_2 A^{}_1 }_{ (\tau^{}_3, t^{}_2, \tau^{}_1 ) } V { \scriptstyle (t^{}_{1} - \tau^{}_1 )}  \nonumber\\
	& + U { \scriptstyle (t^{}_{3} - \tau^{}_3 )} I^{ A^{}_2 A^{}_1 }_{ (\tau^{}_3, t^{}_2, t^{}_1 ) } \nonumber\\
	& =  U { \scriptstyle (t^{}_{3})} A^{}_{2} V^{A^{}_{1}}_{(t^{}_{2}, t^{}_{1})}  \label{eq:Inhomogen_prop_multitime_second} \\
	& +	 U { \scriptstyle (t^{}_{3} - \tau^{}_{3} )} K^{A^{}_{2}}_{(\tau^{}_{3}, \tau^{}_{2})} V^{A^{}_{1}}_{(t^{}_{2} - \tau^{}_{2}, t^{}_{1})} \nonumber \\
	& +  U { \scriptstyle (t^{}_{3} - \tau^{}_{3} )} K^{A^{}_{2}A^{}_{1}}_{(\tau^{}_{3}, t^{}_{2}, \tau^{}_{1})} V { \scriptstyle (t^{}_{1} - \tau^{}_{1})} + U^{}_{(t^{}_{3} - \tau^{}_{3} )} I^{ A^{}_{2} A^{}_{1} }_{(\tau^{}_{3}, t^{}_{2}, t^{}_{1})} . \nonumber
\end{align}
\end{subequations}
A diagrammatic representation of the terms contributing to 
Eqs.~\eqref{eq:Homog_prop_multitime} and \eqref{eq:Inhomogen_prop_multitime} is shown 
in Figs.~\ref{fig:Homog_diag} and \ref{fig:Inhomog_diag}. From 
Fig.~\ref{fig:Homog_diag} we can see that $U^{ A^{}_1 }$ ($ U^{ A^{}_2 A^{}_1 }$) are obtained by constructing all possible combinations between $U$, $A^{}_1$, $K^{A^{}_{1}}_{}$  
(and $A^{}_2$, $K^{A^{}_{2}}_{}$, $K^{A^{}_{2} A^{}_{1}}_{}$), such that $A^{}_1$ (and $A^{}_2$) appear only once in every combination. In addition, the sum of the first two diagrams and the next two diagrams in Fig. \ref{fig:Homog_diag}b give the first and the second term of Eq. \eqref{eq:Homogen_prop_multitime_second}. Also the diagrams contributing to $V^{A^{}_{1}}_{}, V^{A^{}_{2} A^{}_{1}}_{}$ can be obtained by taking all diagrams from Fig. \ref{fig:Homog_diag}a, \ref{fig:Homog_diag}b respectively and replacing the last homogeneous propagator by an inhomogeneous one or by replacing the homogeneous kernels containing $A^{}_{1}$-superscript and the homogeneous propagator on their right side with the corresponding inhomogeneous kernels. The sum of the first three diagrams and the next three diagrams in Fig. \ref{fig:Inhomog_diag}b give the first and the second term of Eq.~\eqref{eq:Inhomogen_prop_multitime_second}.

\subsection{Equations for the multitime kernels}
\label{sec:Eq_for_multitime_kernels}
We will consider the case of having time-independent Liouvillians but the results can be easily generalised to the time-dependent case. Looking at the rules for constructing  diagrams we can conclude that every $N$-time homogeneous propagator ($N>1$) contains an $N$-time homogeneous kernel, the first and last arguments of which are convoluted with $U$. We can always derive an equation for $K^{A^{}_{N-1} \ldots A^{}_{1} }_{}$ by applying a time derivative operator to the first and the last argument of $U^{A^{}_{N-1} \ldots A^{}_1 }_{}$. For the $N=1$ case we just have to take the second time derivative of $U$. A closer look at the time derivatives of a multitime propagators will allow us to cancel a significant amount of terms, which will make the resulting equation numerically more stable. First, we define the following system operators:
\begin{subequations}
\begin{align}
U^{A^{}_{N-1} \ldots A^{}_{1} }_{i \hspace{0.3mm} ( t^{}_{N}, \ldots , t^{}_{1} ) } & = { \rm tr^{}_{B} } \big[ e^{ \mathcal{L} t^{}_{N} }_{} A^{}_{N-1} \ldots e^{ \mathcal{L} t^{}_2 }_{} A^{}_1 e^{ \mathcal{L} t^{}_1 }_{}  \mathcal{L}^{}_{ \bar{S}\bar{B} }  R \big], \\
U^{A^{}_{N-1} \ldots A^{}_{1} }_{f \hspace{0.3mm} ( t^{}_{N}, \ldots , t^{}_{1} ) } & =  { \rm tr^{}_{B} } \big[  \mathcal{L}^{}_{ \bar{S}\bar{B} } e^{ \mathcal{L} t^{}_{N} }_{} A^{}_{N-1} \ldots e^{ \mathcal{L} t^{}_2 }_{} A^{}_1 e^{ \mathcal{L} t^{}_1 }_{} R \big],  \\
U^{A^{}_{N-1} \ldots A^{}_{1} }_{fi \hspace{0.3mm} ( t^{}_{N}, \ldots , t^{}_{1} ) } & =  { \rm tr^{}_{B} } \big[  \mathcal{L}^{}_{ \bar{S}\bar{B} }  e^{ \mathcal{L} t^{}_{N} }_{} A^{}_{N-1} \ldots e^{ \mathcal{L} t^{}_2 }_{} A^{}_1 e^{ \mathcal{L} t^{}_1 }_{}  \mathcal{L}^{}_{ \bar{S}\bar{B} }  R \big],
\end{align}
\end{subequations}
where $\mathcal{L}^{}_{\bar{S}\bar{B}} \equiv \mathcal{L}^{}_{SB} - \langle \mathcal{L}^{}_{SB} \rangle$ and $\mathcal{L}^{}_{\bar{S}} \equiv \mathcal{L}^{}_{S} + \langle \mathcal{L}^{}_{SB} \rangle $. The $i,f$ subscript shows that we have applied $\mathcal{L}^{}_{ \bar{S}\bar{B} }$ at the beginning/end  of the expression before taking the trace. \newline
By use of the fact that
\begin{equation}
\begin{array}{rcl}
\partial^{}_{t^{}_N } U^{A^{}_{N-1} \ldots A^{}_{1} }_{ ( t^{}_{N}, \ldots , t^{}_{1} ) } & = & \mathcal{L}^{}_{\bar{S}} U^{A^{}_{N-1} \ldots A^{}_{1} }_{ ( t^{}_{N}, \ldots , t^{}_{1} ) }  + U^{A^{}_{N-1} \ldots A^{}_{1} }_{f \hspace{0.3mm} ( t^{}_{N}, \ldots , t^{}_{1} ) }  
\end{array}
\end{equation}
we can see that after applying the time derivative w.r.t. $t^{}_N $ at both sides of the equation for $U^{A^{}_{N-1} \ldots A^{}_1 }_{}$, all terms proportional to $\mathcal{L}^{}_{\bar{S}} $ cancel out, such that only $U^{A^{}_{N-1} \ldots A^{}_1 }_{f}$ remains on the left-hand-side of the equation. The same argument is valid also for the case of applying $\partial^{}_{t^{}_1}$ on both sides of the new equation. For the two-time homogeneous propagator $U^{A^{}_{1}}_{}$ we obtain for example the following set of equations:
\begin{subequations}
\allowdisplaybreaks
\begin{align}
U^{A^{}_{1}}_{f \hspace{0.3mm} ( t^{}_{2}, t^{}_{1} ) } & =  U^{}_{ f} { \scriptstyle ( t^{}_{2} )} A^{}_1 U {\scriptstyle (t^{}_1)} \\
	& + U^{}_{f} { \scriptstyle ( t^{}_2 - \tau^{}_2 ) } K^{ A^{}_1 }_{( \tau^{}_2, \tau^{}_1 )} U { \scriptstyle ( t^{}_1 - \tau^{}_1 ) } \nonumber \\
	& + K^{ A^{}_1 }_{} {\scriptstyle (t^{}_2, \tau^{}_1 )} U {\scriptstyle (t^{}_1 - \tau^{}_1 )}, \nonumber \\
U^{A^{}_{1}}_{fi \hspace{0.3mm} ( t^{}_{2}, t^{}_{1} ) } & =  
  	U^{}_{ f } { \scriptstyle ( t^{}_{2} )} A^{}_1 U^{}_{ i } { \scriptstyle (t^{}_1)} \\
  	& + U^{}_{f} { \scriptstyle ( t^{}_2 - \tau^{}_2 ) } K^{ A^{}_1 }_{( \tau^{}_2, \tau^{}_1 )} U^{}_{ i } { \scriptstyle ( t^{}_1 - \tau^{}_1) } \nonumber \\
	& +	U^{}_{f } { \scriptstyle ( t^{}_2 - \tau^{}_2 ) } K^{ A^{}_1 }_{( \tau^{}_2, t^{}_1 )} \nonumber \\
	& + K^{ A^{}_1 }_{( t^{}_2, \tau^{}_1 )} U^{}_{ i } {\scriptstyle ( t^{}_1 - \tau^{}_1 ) } \nonumber \\
	& +	K^{ A^{}_1 }_{( t^{}_2, t^{}_1 )}. \nonumber
\end{align}
\end{subequations}
For the $N=1$ case we derive an equation similar to the Volterra equation of the second kind for $K$, that was defined in \cite{Zhang}. The derivation of an equation for the mutitime inhomogeneous kernels  $I^{A^{}_{N-1} \ldots A^{}_1 }_{}$ can be carried out similarly, the only difference being that only $\partial^{}_{t^{}_N}$ has to be applied on both sides of the equation for $V^{ A^{}_{N-1} \ldots A^{}_1 }_{}$ since it contains always an $N$-time inhomogeneous kernel convoluted on the left side with $U$.

\subsection{Periodically driven systems}
\label{subsec:Per_driv_sys}
	Consider a system which initially is in a unique steady state with its environment. At $t=0$ a periodic force with period $t_P$ is turned on, that is applied at the system. We can simulate numerically this problem by starting from an arbitrary product state $\tilde{\rho}^{}_S \otimes R $, letting it evolve in time until it relaxes to its unique steady state and then turning on the periodic force.
We denote the Liouvillians describing the total system before and after turning on the periodic force by $\mathcal{L}'$ and $\mathcal{L} (t)$, respectively. The state of the system is then given by:
\begin{equation}
\begin{array}{rcl}
\rho^{}_{S } (t) & = & {\rm tr^{}_{B}} \big[  \hat{T} \big( e^{\int^{t}_{0} d\tau \mathcal{L} (\tau) }_{} \big) \mathbb{1} e^{ \mathcal{L}' t^{}_{R} }_{}  \mathcal{P} \tilde{\rho}  \big].
\end{array}
\end{equation}
The time the system, initially prepared in the state $\tilde{\rho} = \tilde{\rho}^{}_{S} \otimes R $, needs to relax to equilibrium will be denoted by $t^{}_{R}$. If we express the unit operator $\mathbb{1}$ as a sum of $\mathcal{P}$ and $\mathcal{Q}$, and apply \textbf{to the last equation} the reduction rules given in Eq.  \eqref{eq:Reduction_rules}, then we obtain the following result:
\begin{align}
\label{eq:EoM_per_driv_sys_in_Eq}
\rho^{}_{S} (t) & =  W ( t , 0, - t^{}_R ) \tilde{\rho}^{}_{S},\\
W ( t^{}_3, t^{}_2, t^{}_1 ) & = \tilde{U} (t^{}_3, t^{}_2 ) U (t^{}_2 - t^{}_1 ) \label{eq:EoM_W_per_driven}\\
+ \mkern-9mu {\textstyle \int\limits^{ t^{}_3 - t^{}_2 }_{ 0 }  \mkern-9mu  d\tau^{}_3 \mkern-9mu  \int\limits^{ t^{}_2 - t^{}_1 }_{ 0 } \mkern-9mu d\tau^{}_1 }
 \tilde{U} & (t^{}_3, t^{}_2  + \tau^{}_{3}) \mathcal{K}  ( \tau^{}_{3}, t^{}_2, \tau^{}_{1} ) U (t^{}_{2} - \tau^{}_{1} - t^{}_1 )  , \nonumber
\end{align}
where we have defined
\begin{align}
\tilde{U} ( t^{}_{2}, t^{}_{1} ) & =  { \rm tr^{}_{B} } \big[ \hat{T} e^{ \int^{ t^{}_2 }_{ t^{}_1 } d\tau \mathcal{L} (\tau) }_{} R \big],  \label{eq:Def_tilde_U} \\
U ( t^{}_{1} ) & =  { \rm tr^{}_{B} } \big[ e^{ \mathcal{L}' t^{}_{1} }_{} R \big],  \\
\mathcal{K} ( t^{}_3, t^{}_2, t^{}_1 ) & = { \rm tr^{}_B } \big[  \mathcal{L}^{}_{SB} (t^{}_{32}) \hat{T}  \Big( e^{ \int^{ t^{}_{32} }_{ t^{}_2 }  d\tau \mathcal{Q} \mathcal{L}  (\tau) }_{} \Big) \times \\
		& \hspace{25.0mm} \times \mathcal{Q} e^{ \mathcal{Q} \mathcal{L}' t^{}_{1} }_{} \mathcal{L}'^{}_{SB} R \big], \nonumber
\end{align}
with $t^{}_{32} \equiv t^{}_3 + t^{}_2 $. Equation \eqref{eq:Def_tilde_U} is just the extension of Eq. \eqref{eq:Def_U} to time dependent Liouvillians. In order to calculate $\tilde{U}$ we also have to use a similar extension of Eq. \eqref{eq:Def_K__with_Lsb}:
\begin{equation}
\label{eq:Def_tilde_K}
\begin{array}{rl}
\tilde{K} (t^{}_{2}, t^{}_{1}) & = {\rm tr^{}_B} \big[ \mathcal{L}^{}_{SB} (t^{}_2) \hat{T} \big(  e^{ \int^{ t^{}_2 }_{ t^{}_1 }  d\tau \mathcal{Q} \mathcal{L} (\tau) }_{}  \big)  \mathcal{Q} \mathcal{L}^{}_{SB} (t^{}_1) R \big].
\end{array}
\end{equation}
The propagators $U$, $\tilde{U}$ describe the time evolution of a system described by $\mathcal{L}', \mathcal{L} (t) $, which is initially in the product state $\rho^{}_{S} \otimes R$ with arbitrary $ \rho^{}_S $. The second argument of $\tilde{U} (t^{}_2, t^{}_1) $ gives the initial phase of the function of the periodic force, while the difference $t^{}_2 - t^{}_1$ is the actual evolution time. It follows that we need to know $\tilde{U} (t^{}_2, t^{}_1) $ only in the range $t^{}_1 \in [0,t^{}_P)$. This property, also valid for $\tilde{K} (t^{}_2, t^{}_1) $, can be seen directly from Eqs. \eqref{eq:Def_tilde_U} and \eqref{eq:Def_tilde_K} and formally reads
\begin{equation}\label{eq:Per_condition_ker_and_prop}
 \mathcal{B} (t^{}_2 + m \cdot t^{}_{P}, t^{}_1 + m \cdot t^{}_{P}) 
 = \mathcal{B} (t^{}_2 , t^{}_1), 
\end{equation}
where $m\in \mathbb{Z}$ and $\mathcal{B} \in \lbrace  \tilde{U}, \tilde{K} \rbrace$.
This fact is important since it allows us to calculate $\tilde{U} (t^{}_2, t^{}_1 ) $ by knowing $\tilde{K} (t^{}_2, t^{}_1) $ only in the range $t^{}_{1} \in [0, t^{}_P)$, $t^{}_2 \in [ t^{}_{1}, f  (t^{}_1) ]$, where $f ( t^{}_1 ) $ is chosen such that $\tilde{K} (t^{}_2 , t^{}_1 ) = 0 $ for $t^{}_2 > f (t^{}_1) $.

In order to better understand how we have to choose the relaxation time $t^{}_R$, and to
explain why this is not the time the system density matrix needs to reach its steady state, we split $\rho^{}_S (t) $ in terms of homogeneous and inhomogeneous propagators as it is done in Eq.  \eqref{eq:Def_rho_s_B}, and solve Eq. \eqref{eq:EoM_per_driv_sys_in_Eq} for the inhomogeneous kernel $\tilde{I}$, which is defined as
\begin{equation}
\begin{array}{rl}
\tilde{I} ( t, 0 ) & = {\rm tr^{}_B} \big[ \mathcal{L}^{}_{SB} ( t ) \hat{T} \big(  e^{ \int^{ t }_{ 0 }  d\tau \mathcal{Q} \mathcal{L} (\tau) }_{}  \big)   \mathcal{Q}  \rho (0) \big].
\end{array}
\end{equation}
	This definition is just the extension of Eq.\eqref{eq:Def_I__with_Lsb} to time dependent Liouvillians. The result is
\begin{equation}
\label{eq:Def_tilde_I_eq_init_state}
\begin{array}{c}
\tilde{I} ( t, 0 ) =  \int^{t^{}_R}_{0} d\tau^{}_1  \mathcal{K} (t, 0, \tau^{}_1 ) U (t^{}_R - \tau^{}_1 ) \tilde{\rho}^{}_S .
\end{array}
\end{equation}
The constraint that the system was initially in its steady state means that $\tilde{I} (t,0) $ is independent of $t^{}_R$ for all $t \geq 0$. If we define by $t'$ the time that the propagator $U$ needs to reach its steady state and by $t''$ the time, where $ \mathcal{K}( t,0,\tau^{}_1) = 0 $   $ \forall \tau^{}_{1} > t''$, then any $t^{}_R > t'+t''$ gives the same result in Eq. \eqref{eq:Def_tilde_I_eq_init_state}. We are free to set $t^{}_R \rightarrow \infty$ and replace $U (t^{}_R - \tau^{}_1) \tilde{\rho}^{}_{S} $ by the steady state of the system $\rho^{}_{S} (0) $, which is not driven by a periodic force. Thus, we obtain the following result:
\begin{equation}
\label{eq:Def_tilde_I_eq_init_state_final_def}
\begin{array}{c}
\tilde{I} ( t, 0 )  = \int^{\infty}_{0} d\tau^{}_1 \mathcal{K} (t, 0, \tau^{}_1) \rho^{}_{S} (0) .
\end{array}
\end{equation}
Finally, we mention that the argumentation of the previous subsection can also be applied to 
Eq.~\eqref{eq:EoM_W_per_driven} in order to obtain an equation for 
$\mathcal{K} (t^{}_3, 0, t^{}_1) $. 

\subsection{Stochastic unravelling method}\label{subsec:Stoch_unravel}

In the following we always assume that the system is coupled to Gaussian environments such that the coupling is linear in the environmental fields. This allows us to integrate out analytically the reservoir degrees of freedom. All (multitime) propagators will then contain the same time nonlocal contribution $u^{}_{NL}$ of the form:
\begin{equation}
\label{eq:Def_U_NL} 
\begin{array}{l}
u^{}_{NL} ( t^{}_2, t^{}_1 ) = \exp \Big[ \sum\limits^{}_{j} 
\int^{ t^{}_2 }_{ t^{}_1 }  d\tau  			\hspace{0.5mm}  i \Upsilon^{ \times }_{j} (\tau) 
\int^{ t^{}_2 }_{ t^{}_1 }  d\tilde{\tau}	\hspace{0.5mm} \theta ( \tau - \tilde{\tau} ) \\
\hspace{8.0mm}	 \times \big(
 g^{}_{N,j} (\tau - \tilde{\tau}) i  \Upsilon^{ \times }_{j}  ( \tilde{\tau} ) +
 g^{}_{D,j} (\tau - \tilde{\tau})    \Upsilon^{ o }_{j} ( \tilde{\tau} ) 
 \big)
 \Big],
\end{array}
\end{equation}
where $\Upsilon$ refers to the system part of the system-bath coupling operators and the index $j$ denotes the different environments to which the system is coupled. We have introduced the superoperator notation  $f^{\times}_{} A \equiv [ f, A ]$ and $f^{o}_{} A \equiv \lbrace f, A \rbrace $. The dissipation and noise kernels $g^{}_{D,j}$ and $g^{}_{N,j}$ contributing to $u^{}_{NL}$ are defined as:
\begin{align}
\allowdisplaybreaks
g^{ }_{ D,j } ( t) & =  \int d\omega J^{}_{j} ( \omega ) \sin( \omega t ),  \label{eq:Def_g_D}\\
g^{ }_{ N,j } ( t) & =  \int d\omega J^{}_{j} ( \omega ) \coth \big( \omega/2T \big) \cos(\omega t) \label{eq:Def_g_N},
\end{align}
where for the spectral densities $J^{}_{j} (\omega) $ we have to use the definition given in Eq. \eqref{eq:Spec_fct_fir_problem} or Eq. \eqref{eq:Spec_fct_sec_problem} depending on the problem we are interested in. 
In the following we will apply the stochastic unravelling method to the case of having a single element in the sum over $j$ and the index $j$ will be omitted.

Our goal is to make the action local in time. The first step to achieve this is to eliminate the $\theta$-function in Eq. \eqref{eq:Def_U_NL}. Since $g^{}_{N} (t)$ is symmetric in $t$, we can replace $ (\theta g^{}_{N}) (t) $ with $\frac{1}{2} g^{}_{N} (t) $. The elimination of $\theta$ from $g^{}_{D}$ requires the introduction of the following Fourier transformation:
\begin{align}
\allowdisplaybreaks
\big( i \theta g^{}_{D} \big) (t-t') 	& = \int \frac{d\omega}{2\pi} f (\omega)  e^{ -i\omega(t-t') }_{}, \label{eq:Def_g_D_b}\\
f  (\omega) & = \frac{\pi}{2} \big( -J (\omega) + J (-\omega) \big)  \\
 & + \frac{i}{2} \int d\varepsilon  \hspace{0.5mm} J (\varepsilon) \bigg( \frac{1}{\varepsilon + \omega}  + \frac{1}{ \varepsilon - \omega }\bigg), \nonumber
\end{align}
where the improper integral over $\varepsilon$ is calculated by the Cauchy principal value method. If $J (\omega) $ and $f (\omega) $ go to zero for large values of $\omega$ we can discretize the integrals over $\omega$ in Eqs. \eqref{eq:Def_g_N} and \eqref{eq:Def_g_D_b} by a finite sum of terms. Then we can make the $u^{}_{NL}$ local in time at the cost of introducing a finite number of Gaussian integrals and Eq. \eqref{eq:Def_U_NL} transforms to:
\begin{align}
\label{eq:U_NL_unravelled}
& \hspace{1.0mm} \prod\limits^{}_{\nu} 
\int \frac{ dx^{}_{\nu} d\tilde{x}^{}_{ \nu } }{2\pi} 
\exp\Big[ \frac{- x^{2}_{\nu} - \tilde{x}^{2}_{ \nu } }{2} \Big] \nonumber \\
& {\scriptstyle \times} \prod\limits^{}_{\tilde{\nu}}
\int \frac{ dy^{}_{ \tilde{\nu} } d\tilde{y}^{}_{ \tilde{\nu} } }{2\pi} 
\exp \Big[ \frac{ - y^{2}_{\tilde{\nu}} - \tilde{y}^{2}_{ \tilde{\nu} }}{2} \Big] \nonumber \\
& {\scriptstyle \times} \exp \Big[  \int^{t^{}_2 }_{ t^{}_1 }  d\tau \big( 
\chi (\tau; y, \tilde{y} ) \Upsilon^{\times}_{} (\tau)  + \tilde{\chi} (\tau; y, \tilde{y} ) \Upsilon^{o}_{ }  (\tau)
 \big) \Big] \nonumber \\
& {\scriptstyle \times} \exp \Big[  \int^{t^{}_2 }_{ t^{}_1 }  d\tau  
\xi (\tau; x, \tilde{x} ) i \Upsilon^{\times}_{}  (\tau)  \Big].
\end{align}
The functions $ \xi, \chi, \tilde{\chi}$ are given by:
\begin{align}
\xi (t; x, \tilde{x} ) & = \sum\limits^{\nu_{c}}_{\nu=0} 
\Big[ a^{}_{\nu} J (\omega^{}_{\nu} ) \coth \Big(\frac{\omega^{}_{\nu}}{2T} \Big) \Big]^{1/2} \nonumber \\
 & \hspace{15.0mm} {\scriptstyle \times} \Big[  \cos( \omega^{}_{\nu} t) x^{}_{\nu} + \sin ( \omega^{}_{\nu} t)\tilde{x}^{}_{\nu}  \Big], \\
\chi (t; y,\tilde{y}) & = \sum\limits^{\tilde{\nu}_{c}}_{\nu=-\tilde{\nu}_{c}} 
\Big[  \frac{\Delta \nu}{2} f (\omega^{}_{\nu}) \Big]^{1/2}_{} e^{ -i\omega^{}_{\nu} t }_{}  \big[ y^{}_{\nu} + i \tilde{y}^{}_{\nu} \big], \\
\tilde{\chi} (t; y,\tilde{y} ) & = \sum\limits^{\tilde{\nu}_{c}}_{\nu=-\tilde{\nu}_{c}} 
\Big[  \frac{\Delta \nu}{2} f  (\omega^{}_{\nu}) \Big]^{1/2}_{} e^{ +i\omega^{}_{\nu} t }_{}  \big[ y^{}_{\nu} - i \tilde{y}^{}_{\nu} \big]
\end{align}
with $\omega^{}_{\nu} = 2\pi\nu  \hspace{0.1mm} \Delta \nu$, $a^{}_{\nu} = 2 \pi \hspace{0.1mm} \Delta \nu$ for $\nu > 0$,  
 $a^{}_{0} = \pi \hspace{0.1mm} \Delta \nu$ and $\nu^{}_c, \tilde{\nu}^{}_c$ are properly chosen cutoffs of $J (\omega) $ and $f (\omega) $, respectively. By the use of Monte-Carlo integration techniques for the calculation of the Gaussian integrals we can interpret $x^{}_{\nu},\tilde{x}^{}_{\nu}, y^{}_{\nu}, \tilde{y}^{}_{\nu}$ as normally distributed random variables and $\xi (t) ,\chi (t) ,\tilde{\chi} (t) $ as Gaussian random variables with zero mean and the following correlation functions:
\begin{align}
 \langle \xi (t) \xi (t') \rangle & = g^{}_{N} (t-t'), \\
 \langle \chi (t) \tilde{\chi} (t') \rangle & = i  \theta  (t-t') g^{}_{D} (t-t'). 
\end{align}
All other correlations are equal to zero. This stochastic unravelling of $u^{}_{NL}$ is just a specific realisation of the general idea explained in \cite{Stockburger_02}.\newline

The calculation of an arbitrary multitime propagator $U^{ A^{}_{N-1} \ldots A^{}_{1} }_{}$ is carried out by replacing $u^{}_{NL}$ by the last two lines of Eq. \eqref{eq:U_NL_unravelled} in the definition of $U^{ A^{}_{N-1} \ldots A^{}_{1} }_{}$ and averaging the result over a large enough number of realisations of the normally distributed random variables $\lbrace x^{}_{\nu}, \tilde{x}^{}_{\nu}, y^{}_{\nu}, \tilde{y}^{}_{\nu} \rbrace$. $U^{ A^{}_{N-1} \ldots A^{}_{1} }_{m}$ ($ m \in \lbrace f,i,fi \rbrace$) is calculated by multiplying every realisation of $U^{ A^{}_{N-1} \ldots A^{}_{1} }_{ }$ on the left or/and on the right by $F (t^{}_{N\ldots1}) $ and $ F (0) $, respectively, where $F$ is given by
\begin{equation}
\begin{array}{c}
F (t) = ( i \xi (t) + \chi (t) ) \Upsilon^{\times}_{} +  \tilde{\chi} (t) \Upsilon^{o}_{}.
\end{array}
\end{equation}

\section{Results }
\label{sec:Results}
\subsection{Periodically driven system initially in equilibrium with its environment}
\label{subsec:Results_per_driven_sys}
We apply the formalism derived in Sec.~\ref{sec:II} to the problem of a classically driven two-level system coupled to a bosonic environment \cite{Leggett1987}. The system is described by the following Hamiltonian:
\begin{eqnarray} \label{eq:Hamiltonian_for_Per_driving}
H &=&  \frac{\Delta}{2} \sigma^{}_{z} + \frac{J}{2}\sigma^{}_{x} + \frac{\varepsilon  (t) }{2} \sigma^{}_{z} \nonumber \\
  && + \sum^{}_{k} \big[ \omega^{}_k b^{\dagger}_{k} b^{}_k + \lambda^{}_k ( b^{\dagger}_k + b^{}_k )\sigma^{}_z   \big],
\end{eqnarray}
where $b^{\dagger}_k,b^{}_k$ are the bosonic creation and annihilation operators for the modes $k$ with frequency $\omega^{}_k$, and $\lambda^{}_k$ describes the strength of the interaction of the two-level system with its environment. The Pauli spin-$\frac{1}{2}$ operators are denoted by $\sigma^{}_{j}$ $(j \in \lbrace x,y,z \rbrace)$, the energy distance between the two levels of the spin is $\Delta$ and the classical driving force applied to the system is given by $\varepsilon (t) = V^{}_0 \sin (\Omega t )  $. We set $R = \rho^{eq}_{B} = \frac{1}{N}\prod^{}_k \exp [ -\beta \omega^{}_k b^{\dagger}_{k}  b^{}_k]$ with $\beta = 1/T$ ($k^{}_B = 1 = \hbar $), which allows us to describe the effect of the environment on the two-level system completely by the spectral density:
\begin{equation}
\label{eq:Spec_fct_fir_problem}
\begin{array}{rcl}
J ( \omega ) & = & \frac{2\lambda^{}_{} }{ \pi } \frac{\gamma^{}_{} \omega^2_u \omega }{  ( \omega^2_u - \omega^2_{} )^2_{}  - \gamma^2_{} \omega^2_{} }, \hspace{3.0mm} \omega > 0.
\end{array}
\end{equation}
This form of the spectral density together with the replacement of $\coth(x) $ by $ \frac{1}{x} + \frac{2x}{x^2 + \pi^2} + \frac{2x}{x^2 + 4\pi^2}$ allow us to compare our results with the HEOM method since the dissipation and noise kernels $g^{}_{D}$ and $g^{}_{N}$, defined in 
Eqs.~\eqref{eq:Def_g_D} and \eqref{eq:Def_g_N},  can be expressed by a finite sum of exponentially decaying functions.

We choose the following parameters (measured in units of $\Delta$): $J=0.7$, $V^{}_0 =0.5$, $\lambda^{}_{} = 0.05$, $\gamma^{}_{} = 1.3$, $\omega^{}_u = 0.7$, $\Omega = \pi/2$ and $T=0.35$. By use of the stochastic unravelling method we simulate $U^{}_{l} (t) $ ($l \in \lbrace i, fi \rbrace$) in the range $t\in [0,12]$, $ \tilde{U}^{}_{l} (t^{}_2, t^{}_1) $ ($l \in \lbrace i, f, fi \rbrace$) in the range $t^{}_1 \in [ 0, t^{}_{P})$, $t^{}_2 \in [t^{}_1, t^{}_1 + 12]$ and $W^{}_{fi} (t^{}_3, 0, -t^{}_1) $ for all $(t^{}_{3}, t^{}_1)$ which fulfil the constraint $t^{}_1 + t^{}_3 \leq 12 $. From the equations for the kernels
\begin{align}
K(t) & = U^{}_{fi}(t) - { \textstyle \int^t_{0} } d\tau K(\tau) U^{}_{i} (t-\tau),\\
\tilde{K}(t^{}_2,t^{}_1) & = U^{}_{fi}(t^{}_2,t^{}_1) - { \textstyle \int^{ t_2}_{ t_1} } d\tau \tilde{K}(t^{}_2, \tau) \tilde{U}^{}_{i} ( \tau, t^{}_1),\\
\mathcal{K}(t^{}_3,0, t^{}_1) & = W^{}_{fi}(t^{}_3,0,-t^{}_1)
-\tilde{U}^{}_{f}(t^{}_3,0) U^{}_{i}(t^{}_1) \\
 & - {\textstyle \int^{ t^{}_3 }_{0} d\tau \int^{t^{}_1}_{0} d\tilde{\tau} } \tilde{U}^{}_{f} (t^{}_3,\tau) \mathcal{K}(\tau,0,\tilde{\tau}) U^{}_i ( t^{}_1-\tilde{\tau}) \nonumber \\
 & - {\textstyle \int^{t^{}_3}_{0} d\tau } \tilde{U}^{}_{f} (t^{}_3,\tau) \mathcal{K}(\tau , 0, t^{}_1) \nonumber \\
 & - {\textstyle \int^{t^{}_1}_{0} d\tau } \mathcal{K} (t^{}_3,0,\tau) U^{}_{i} (t^{}_1 - \tau) \nonumber 
\end{align}
we obtain $K (t) , \tilde{K} (t^{}_2, t^{}_1) , \mathcal{K} ( t^{}_3, 0, t^{}_1 ) $  for the same range of times as $U^{}_l (t) , \tilde{U}^{}_l (t^{}_2, t^{}_1) , W^{}_{fi}  (t^{}_3, 0, -t^{}_1) $ respectively ($l\in \lbrace i,f,fi \rbrace$). Since $K (t) ,\mathcal{K} (t^{}_3, 0, t^{}_1) $ are equal to zero outside this range and $\tilde{K} (t^{}_2, t^{}_1)  = 0 $ for $t^{}_1 \in [0, t^{}_P ), t^{}_{2}>t^{}_1 + 12 $, and the periodicity condition in Eq. \eqref{eq:Per_condition_ker_and_prop}, we are able to calculate $U (t), \tilde{U} (t^{}_2, t^{}_1) , W (t^{}_3, 0, -t^{}_1 ) $ for arbitrary $t$, $(t^{}_2, t^{}_1)$ and $(t^{}_3, t^{}_1 )$. The equations for $U,\tilde{U},W$ are obtained by the use of the reduction rules \eqref{eq:Reduction_rules}. Those for $U$ and $W$ are given explicitly in \eqref{eq:Def_U_wieder} and \eqref{eq:EoM_W_per_driven} and the equation for $\tilde{U}$ is given by
\begin{align}
\tilde{U}(t^{}_2,t^{}_1) & = \tilde{u}^{}_{\bar{S}}(t^{}_2, t^{}_1) \\
 & + {\textstyle \int^{t^{}_2}_{t^{}_1} d\tau \int^{\tau}_{t^{}_1} d\tilde{\tau} } \tilde{u}^{}_{\bar{S}} (t^{}_2, \tau) \tilde{K}(\tau, \tilde{\tau}) \tilde{U}(\tilde{\tau}, t^{}_1)  . \nonumber
\end{align}

\begin{figure}
  \centering
  \includegraphics[width=.8\linewidth]{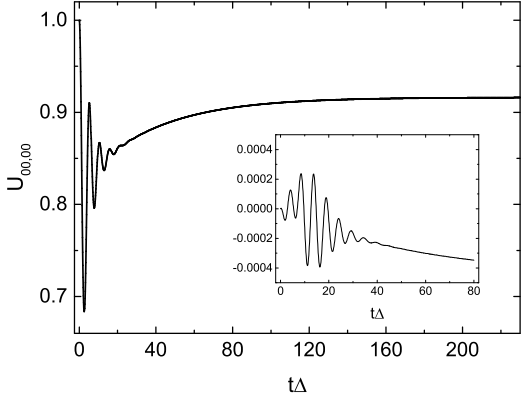}
  \caption{	The time evolution of $U^{}_{00,00} (t) $. Inset: The difference of $U^{}_{00,00} (t) $ 
  and the exact evolution.}
	\label{fig:Graph_W}
\end{figure}

\begin{figure}
	\centering
  	\includegraphics[width=.8\linewidth]{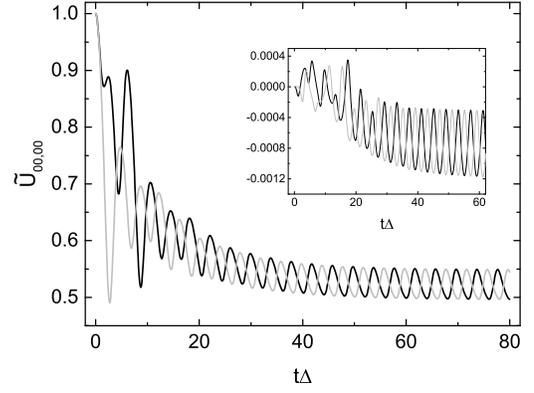}
  	\caption{	The time evolution of $\tilde{U}_{00,00} (t + t^{}_1 , t^{}_1) $ for $t^{}_1 \Delta = 0$ (black line) and $t^{}_1 \Delta = 2$ (gray line). Inset: The difference of $ \tilde{U}^{}_{00,00} (t + t^{}_1 , t^{}_1) $ and the result obtained by use of the HEOM approach.}
  	\label{fig:Graph_U_tilde}
\end{figure}

\begin{figure}
	\centering
	\includegraphics[width=.8\linewidth]{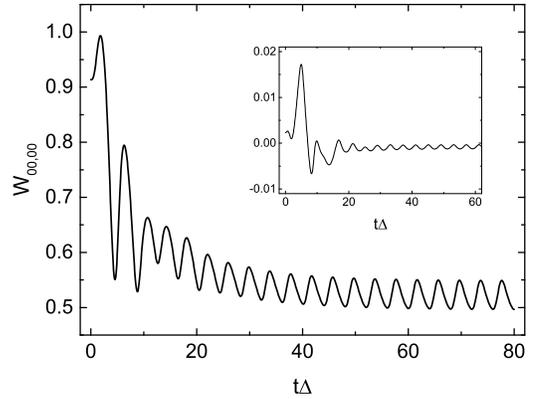}
	\caption{	The time evolution of $W^{}_{00,00} (t,0,-t^{}_R) $ for $t^{}_R \Delta = 240 $. Inset: The
	 difference of $W^{}_{00,00}$ and the exact solution.}
	\label{fig:Graph_U1}
\end{figure}

In the following we denote the up- and down-state of the two-level system by $ |1\rangle $ and $ |0\rangle $. This means that $\sigma^{}_{z} = |1\rangle \langle 1| - | 0 \rangle \langle 0 |$, $ \sigma^{}_{x} = |1\rangle \langle 0| + | 0 \rangle \langle 1 | $. An element of some system superoperator $M$ will be denoted by
\begin{equation}
\begin{array}{c}
M^{}_{ij,kl} = {\rm tr^{}_S} \big[ ( |i \rangle \langle j| )^{\dagger}_{} M |k \rangle \langle l|  \big], \hspace{3.0mm} i,j \in  \lbrace 0,1 \rbrace .
\end{array}
\end{equation}
If we plot all elements of $U \scriptsize{(t)}$ we will see that they become constant for  $t > 200 $. In addition, the elements of $U$ in its steady state obey the following relations:
\begin{eqnarray}
\label{eq:Relations_U_eq}
 & U^{}_{00,00} = U^{}_{00,11} = 1 - U^{}_{11,00} = 1 - U^{}_{11,11}, \label{eq:Relations_U_eq_1} \\
 & U^{}_{01,00} = U^{}_{01,11} =  U^{*}_{10,00} = U^{*}_{10,11} \label{eq:Relations_U_eq_2}
\end{eqnarray}
and all other elements of $U$ are equal to zero. This assures that for every initial system density matrix $\tilde{\rho}^{}_{S}$ the final steady state $U  (t) \tilde{\rho}^{}_S $ $(t>200)$ is the same. Taking into account that $\mathcal{K} (t^{}_3, 0 , t^{}_1) = 0 $ for $t^{}_1 > 12 $ we set $t^{}_R=240$. In Fig. \ref{fig:Graph_W} we have plotted the time evolution of $ U^{}_{00,00} ( t ) $, which represents the occupation of the lower energy site given that the system was initially in $\rho^{}_S (0)  = |0 \rangle \langle 0| \otimes \rho^{eq}_{B}$. The difference between $U^{}_{00,00} (t) $ and the exact solution (obtained by the use of the HEOM approach) origins mainly from the large time step $(0.04)$ used in the calculation of $K$ and $U$.\newline

In Fig. \ref{fig:Graph_U_tilde} we can see the time evolution of $\tilde{U}^{}_{00,00} (t + t_1, t_1 ) $ for $t^{}_1 = 0$ and $t^{}_1 = 2$ which corresponds to the case of having a driving force of the form  $\varepsilon (t) = \sin(\Omega t)$ and $\varepsilon  (t) = \sin(\Omega t + \pi)$ respectively. For large enough $t $ $\tilde{U} (t + t_1,t_1) $ satisfies Eq. \eqref{eq:Relations_U_eq_1},\eqref{eq:Relations_U_eq_2} and also has the property that $\tilde{U} (t^{}_2, t^{}_1) =  \tilde{U} (t^{}_2 , 0) $. This means that the steady state of a system being initially in $\rho^{}_S (0) \otimes \rho^{eq}_{B} $ (and being evolved with $H$ given in Eq. \eqref{eq:Hamiltonian_for_Per_driving} with $V^{}_0 =0.5$) does not depend on $\rho^{}_{S} (0) $ but only on the initial phase of the driving force.\newline


The time evolution of $W^{}_{00,00} (t, 0, -t^{}_R) $ is given in Fig.~\ref{fig:Graph_U1}. From the inset in the figure we can see that the error increases by an order of magnitude in comparison to the previous two cases. The growth of the error origins from the second line of Eq. \eqref{eq:EoM_W_per_driven}, where $\mathcal{K}$ is convoluted with the functions $U,\tilde{U}$, which , as shown in Fig. \ref{fig:Graph_W},\ref{fig:Graph_U_tilde}, deviate from the exact result. 
Even in this case the relative error remains below $2 \% $ at short time scales and below $1 \% $ at long time scales.

\subsection{2D-spectra of a donor-acceptor model }
\label{subsec:2D_spec_fct}

As a second example we calculate the 2D-spectra of a system composed of a single donor and acceptor, each of them coupled to a different phononic bath. We reduce the description to the zero- and single exciton manifold. The total Hamiltonian is given by:
\begin{equation}
\label{eq:Hamiltonian_for_2D_spec}
\begin{array}{rcl}
H & = & \sum\limits^{}_{j=1,2} (\Delta^{}_j + \Delta^{}_{j,r}) | j \rangle \langle j | + J( | 1 \rangle \langle 2 | + | 2 \rangle \langle 1 | ) \\
  & + & \sum\limits^{}_{j=1,2} \sum\limits^{}_{k} \big[ \omega^{}_{jk} b^{\dagger}_{jk} b^{}_{jk} + \lambda^{}_{jk}( b^{\dagger}_{jk} + b^{ }_{jk} ) | j \rangle \langle j |  \big],
\end{array}
\end{equation}
where $\Delta^{}_{1}, \Delta^{}_{2}$ are the energy levels of the states where only the donor $ | 1 \rangle $ or the acceptor $ | 2 \rangle $ are excited. The energy level of the ground state $ | 0 \rangle $  is set to zero. The reorganisation energies $\Delta^{}_{1,r},\Delta^{}_{2,r}$ are defined as
\begin{equation}
\begin{array}{rl}
\Delta^{}_{j,r} & = \int d\omega J^{}_{j} ( \omega ) / \omega, \hspace{6.0mm} j\in \lbrace 1,2 \rbrace.
\end{array}
\end{equation}
The second line of Eq. \eqref{eq:Hamiltonian_for_2D_spec} describes the reservoir and the system-reservoir interaction in a similar way as it is done in  Eq.~\eqref{eq:Hamiltonian_for_Per_driving}. The spectral densities describing the effect of both environments on the system are given by:
\begin{equation}
\label{eq:Spec_fct_sec_problem}
\begin{array}{rcl}
J^{}_j ( \omega ) & = & \frac{2\lambda^{}_{j} }{ \pi } \frac{\gamma^{}_{j} \omega^2_{u,j} \omega }{  ( \omega^2_{u,j} - \omega^2_{} )^2_{}  - \gamma^2_{j} \omega^2_{u,j} }, \hspace{3.0mm} \omega > 0.
\end{array}
\end{equation}
This means that we have already assumed that the initial state of the system is of the form 
$\rho^{}_{S} (0) \otimes \rho^{eq}_B $, where $\rho^{eq}_B =  \frac{1}{\mathcal{N}} \prod^{}_{j} \prod^{}_{k} \exp [ -\beta \omega^{}_{jk} b^{ \dagger }_{jk} b^{}_{jk} ]$ with normalization constant $\mathcal{N}$.

The 2D-spectra $I ( \Omega^{}_3, t^{}_2, \Omega^{}_1 ) $ is defined as a double Fourier transform of the rephasing $R^{}_{rp}$ and nonrephasing $R^{}_{nr}$ contributions to the 
third-order optical response function \cite{Mukamel_1995}:
\begin{align}
\allowdisplaybreaks
I ( \Omega^{}_3, t^{}_2, \Omega^{}_1 ) & =
{\Re}
{\textstyle \int\limits_{0}^{\infty} \mkern-2mu dt^{}_{1} \int\limits^{\infty}_{0} \mkern-2mu dt^{}_{3} }
e^{i( \Omega^{}_{1}t^{}_{1} + \Omega^{}_{3}t^{}_{3} )}_{} R^{}_{nr}  (t^{}_{3}, t^{}_{2}, t^{}_{1} )  + \nonumber \\
 & +
{\Re}
{\textstyle \int\limits_{0}^{\infty} \mkern-2mu dt^{}_{1} \int\limits^{\infty}_{0} \mkern-2mu dt^{}_{3} }
e^{i( -\Omega^{}_{1}t^{}_{1} + \Omega^{}_{3}t^{}_{3} )}_{} R^{}_{rp}  (t^{}_{3}, t^{}_{2}, t^{}_{1} ) ,  \\
R^{}_{rp} (t^{}_{3}, t^{}_{2}, t^{}_{1} ) & = 
 {\rm tr^{}_S}\big[\mu^{ }_{L}  U^{ \mu^{\times}_{R} \mu^{\times}_{R} }_{ (t^{}_3, t^{}_2, t^{}_1) }  \mu^{\times}_{L}  \rho^{}_S (0) \big], \\
R^{}_{nr} (t^{}_{3}, t^{}_{2}, t^{}_{1} ) & =
 {\rm tr^{}_S}\big[\mu^{ }_{L}  U^{ \mu^{\times}_{R} \mu^{\times}_{L} }_{ (t^{}_3, t^{}_2, t^{}_1) }  \mu^{\times}_{R}  \rho^{}_S (0) \big].
\end{align}
The operators $\mu^{}_{L},\mu^{}_{R}$ are contributions to the total dipole operator $\mu = \mu^{}_L + \mu^{}_R $, where $\mu^{}_L = \mu^{}_{1} |0 \rangle \langle 1 | + \mu^{}_{2} |0 \rangle \langle 2 | $ and $\mu^{}_R = \mu^{}_{1} |1 \rangle \langle 0 | + \mu^{}_{2} |2 \rangle \langle 0 | $. The system is initially in its ground state $\rho^{}_S {\scriptsize (0)} = | 0 \rangle \langle 0 |$.

In order to calculate $U^{ \mu^{\times}_{R} \mu^{\times}_{R} }_{}$ and $U^{ \mu^{\times}_{R} \mu^{\times}_{L} }_{}$ we use  Eq.~\eqref{eq:Homogen_prop_multitime_second}. The terms on the right-hand side contain at most two time integrals, which substantially simplifies the
numerical simulations. From Eqs.~\eqref{eq:Def_U_wieder}, \eqref{eq:Homogen_prop_multitime_first}, \eqref{eq:Homogen_prop_multitime_second} we can derive equations for $K$, ($K^{ \mu^{\times}_{L} }_{}$, $K^{ \mu^{\times}_{R} }_{}$) and ($K^{ \mu^{\times}_{R}  \mu^{\times}_{R}}_{}$, $K^{ \mu^{\times}_{R}  \mu^{\times}_{L}}_{}$) respectively. They also contain only terms with at most two time integrals. The equations are solved for a finite time interval defined by the parameters $\tilde{t}, \tilde{t}^{}_{\mu}, \tilde{t}^{}_{\mu\mu}$ as follows: $K (t)$ is calculated for $t<\tilde{t}$, $K^{ \mu^{\times}_{L} }_{ (t^{}_2,t^{}_1) }$ and  $K^{ \mu^{\times}_{R} }_{ (t^{}_2,t^{}_1) } $  - for $(t^{}_2, t^{}_1)$, which fulfil the condition $t^{}_2 + t^{}_1 < \tilde{t}^{}_{\mu}$, $K^{ \mu^{\times}_{R}  \mu^{\times}_{R}}_{ (t^{}_3,t^{}_2,t^{}_1) }  $, $K^{ \mu^{\times}_{R}  \mu^{\times}_{L}}_{ (t^{}_3,t^{}_2,t^{}_1) } $ - for $(t^{}_3,t^{}_2,t^{}_1)$ which fulfil the condition $t^{}_1 + t^{}_2 + t^{}_3 < \tilde{t}^{}_{\mu \mu}$. The parameters $\tilde{t},\tilde{t}^{}_{\mu},\tilde{t}^{}_{\mu\mu}$ are chosen such that the kernels are zero outside the corresponding range. After solving the equations for the kernels we calculate $U$, which is used in the calculation of $U^{\mu^{\times}_{R}}_{}$, $U^{\mu^{\times}_{L}}_{}$.
Finally, the three propagators are used in the  calculation of $U^{ \mu^{\times}_{R} \mu^{\times}_{R} }_{}$, $U^{ \mu^{\times}_{R} \mu^{\times}_{L} }_{}$.

\begin{figure}
  \centering
  \includegraphics[width=0.96\linewidth]{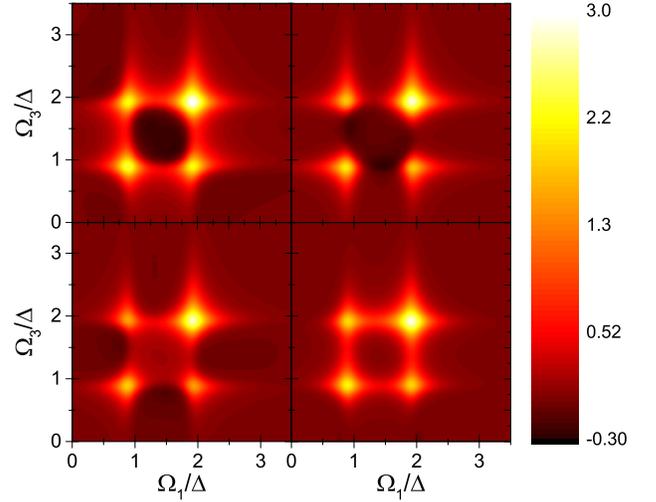}
  \caption{(Color online) Simulated 2D-spectra of the donor-acceptor model at $t\Delta=0$, $t\Delta=1.5$, $t\Delta=2.5$, $t\Delta=5$ (from left to right and from top to bottom). We have used an
arcsinh scaling for the color bar as in Ref. \cite{Engel_2007}.}
	\label{fig:2D_spec}
\end{figure}

\begin{figure}
  \centering
  \includegraphics[width=0.85\linewidth]{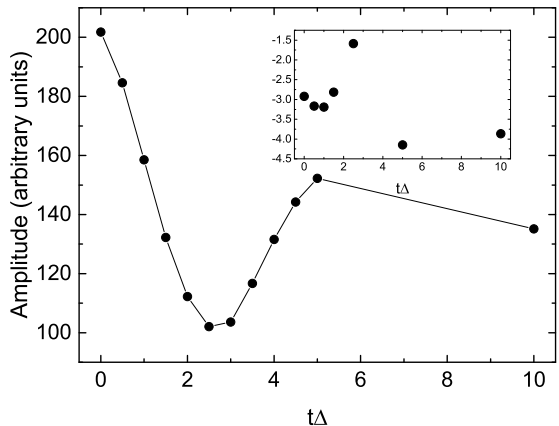}
  \caption{	The time evolution of the hight of the upper left peak of the 2D-spectra. Inset: The 
  difference of the amplitude and the exact result.}
	\label{fig:2D_spec_off_diagonal}
\end{figure}

We work with the following parameters (measured in units of $\Delta \equiv \Delta^{}_2 + \Delta^{}_{2,r}$): $\Delta^{}_{1} =1.9, \Delta^{}_{2} = 0.9, J=0.1, \lambda^{}_1 = \lambda^{}_2 = 0.1, \omega^{}_1 = \omega^{}_2 = 1.4, \gamma^{}_1  = \gamma^{}_2 = 2.6 $. The reorganisation energies are $\Delta^{}_{1,r} = \Delta^{}_{2,r} =0.1 $  and the memory kernels are calculated in the range defined by $\tilde{t} = 5$, $\tilde{t}^{}_{\mu} = \tilde{t}^{}_{\mu\mu} = 4$. The 2D-spectra of the system are plotted for four different waiting times $t^{}_2$ in Fig. \ref{fig:2D_spec}. Since the plots obtained by a direct use of the HEOM look the same as those of Fig. \ref{fig:2D_spec}, we compare the time evolution of the off-diagonal peaks in Fig. \ref{fig:2D_spec_off_diagonal} in order to obtain information about the quantitative accuracy of the method. The relative error is below $3 \%$, which in this case is accurate enough to reproduce all main features of the 2D-spectra.

In general, the error depends on the number of time arguments of the multitime propagator, from which the observable is derived. The higher this number is, the more integro-differential equations for the kernels have to be solved. Since the calculation of an $n$-time propagator requires not only the knowledge of the kernels but also of a set of $m$-time propagators ($m<n$), the error accumulates.

\section{Concluding remarks}
\label{sec:Conclusion}

In this paper we have extended the Nakajima-Zwanzig projection operator technique to the calculation of multitime correlation functions, which required the introduction of multitime kernels. The applicability of the theory was demonstrated by simulating the time evolution of a driven two level system being initially in equilibrium with its environment, and by determining the 2D-spectra of a donor-acceptor model. 
It is important to mention that we have considered systems with environmental spectral densities of the form of Eq.~\eqref{eq:Spec_fct_sec_problem} because the HEOM approach is well  suited for such problems. If we work with an environment whose dissipation and noise kernels can not be approximated by a finite number of exponentially decaying functions, then the combination of stochastic unravelling and the equations for multitime kernels should become the preferable approach since its complexity, in contrast to the HEOM approach, will not increase as long as the kernels decay sufficiently fast. 

In the examples of Sec.~ \ref{sec:Results} we have always assumed that all kernels are nonzero only for a finite range of times. However, this assumption is of course not fulfilled for all models of interest. As a trivial counterexample we can consider a spin-boson model with Hamiltonian
\begin{equation}
\begin{array}{c}
H =  \frac{\Delta}{2}\sigma^{}_{z} + \omega b^{\dagger}_{}b + \lambda ( b^{\dagger}_{} + b ) \sigma^{}_x.
\end{array}
\end{equation}
The spectral density of the environment contains a $\delta$-peak which results in non-decaying multitime kernels. In general, we expect that the multitime kernels will decay sufficiently fast to zero if the spectral density of the environment is smooth enough. In cases where the stochastic unravelling method fails and the system-bath coupling $\lambda{}_{SB}$ is weak enough, we can still try to approximate all multitime kernels by expanding them in powers of $\lambda{}_{SB}$ and taking only the first few terms into account.

Besides the slow decay of the memory kernel, another problem for our approach could be the size of the system. For a system Hilbert space of dimension $\mathcal{N}$ we have to work with kernels and propagators which are represented by $\mathcal{N}^2_{} \times \mathcal{N}^{2}_{}$ matrices. This has to be compared with the $\mathcal{N}^2_{}$-dependence of the HEOM method on the system size.
	
Additional problems arise from the fact that the calculation of an $m$-time propagator in its full time domain requires the knowledge of all $m'$-time propagators and $m''$-time kernels
$(m'<m, m''\leq m)$ in their full time domain, which leads to accumulation of the numerical error by an increase of $m$.
	 
If we want to calculate an $(m+n)$-time propagator in the time domain, where $n$ of its arguments are fixed, we can not guarantee that we will have to know only a set of kernels/propagators, whose time domain is at most $m$-dimensional. In the first example that we have considered in 
Sec.~\ref{subsec:Results_per_driven_sys} we have fixed the last two arguments of $W(t^{}_3, t^{}_2, t^{}_1)$ to $(t^{}_2, t^{}_1)=(0,-t^{}_R)$. But for the calculation of $W$ in its one-dimensional time domain we needed the pairs $\mathcal{K},U$ and $\tilde{K}, \tilde{U}$, whose time domains were one- and two-dimensional, respectively. On the other hand, in the second example in 
Sec.~\ref{subsec:2D_spec_fct} we have fixed the second argument of $I$ and $U^{ \mu^{\times}_{R} \mu^{\times}_{R} }_{}, U^{ \mu^{\times}_{R} \mu^{\times}_{L} }_{}$. For the calculation of $U^{ \mu^{\times}_{R} \mu^{\times}_{R} }_{}, U^{ \mu^{\times}_{R} \mu^{\times}_{L} }_{}$ we used kernels/propagators whose time domains were at most two-dimensional.
	
In summary, we have presented a method for the calculation of MTCFs of systems which span finite Hilbert spaces. In the first step we calculate the kernels via a set of equations. The input information can be obtained via a modification of the HEOM method or via a stochastic unravelling method. In the second step the kernels are used for the calculation of MTCFs by use of equations which can be derived by a few simple reduction rules. Thus, the main advantage of the present method is that it can be applied to problems, where HEOM does not perform well, as long as the system is sufficiently small and the memory kernels decay sufficiently fast.


\acknowledgments

HPB acknowledges support from the EU Collaborative Project QuProCS (Grant Agreement
641277).


\begin{thebibliography}{}

\bibitem{TheWork}
H.-P. Breuer and F. Petruccione, \textit{The Theory of Open Quantum Systems}, (Oxford 
University Press, Oxford, 2002).

\bibitem{Vulto_1999}
	S. I. E. Vulto, M. A. de Baat, S. Neerken, F. R.  Nowak, H. van Amerongen, J. Amesz, 
	T. J. Aartsma,
	J. Phys. Chem. B \textbf{103}, 8153 (1999).

\bibitem{Jang_2002}
	S. Jang, Y. J. Jung, R. J. Silbey,
	Chem. Phys. \textbf{275}, 319 (2002).	
	
\bibitem{Jang_2004}
	S. Jang, M. D. Newton, R. J. Silbey,
	Phys. Rev. Lett. \textbf{92}, 218301 (2004).

\bibitem{Mohseni_2008}
	M. Mohseni, P. Rebentrost, S. Lloyd, A. Aspuru-Guzik,
	J. Chem. Phys. \textbf{129}, 174106 (2008).
	
\bibitem{Castro_2008}
	A. Olaya-Castro, C. F. Lee, F. F. Olsen, N. F. Johnson,
	Phys. Rev. B \textbf{78}, 085115 (2008).

\bibitem{Braig_2003}
	S. Braig, K. Flensberg,
	Phys. Rev. B \textbf{68}, 205324 (2003).

\bibitem{Mitra_2004}
	A. Mitra, I. Aleiner, and A. J. Millis,
	Phys. Rev. B \textbf{69}, 245302 (2004).

\bibitem{Koch_2005}
	J. Koch, F. von Oppen,
	Phys. Rev. Lett \textbf{94}, 206804 (2005).

\bibitem{Haenggi_2013}
	J. Thingna, J. Wang, and P. H\"anggi,
	Phys. Rev. E \textbf{88}, 052127 (2013).	

	
\bibitem{Aligia_2006}
	A. A. Aligia,
	Phys. Rev. B \textbf{74}, 155125 (2006).

\bibitem{Monreal_2010}
	R. C. Monreal, F. Flores, and A. Martin-Rodero,
	Phys. Rev. B \textbf{82}, 235412 (2010).

\bibitem{Nakajima}
S. Nakajima, Progr. Theor. Phys. \textbf{20}, 948 (1958).

\bibitem{Zwanzig}
R. Zwanzig, J. Chem. Phys. \textbf{33}, 1338 (1960).

\bibitem{Shibata}
F. Shibata, Y. Takahashi, and N. Hashitsume, 
J. Stat. Phys. \textbf{17}, 171 (1977).

			
\bibitem{Daley_2004}
	A. J. Daley, C. Kollath, U. Schollw\"ock, G. Vidal,
	 J. Stat. Mech.: Theor. Exp. P04005 (2004).

\bibitem{White_2004}
	S. R. White, A. E. Feiguin,
	Phys. Rev. Lett. \textbf{93}, 076401 (2004).

\bibitem{Schmitteckert_2004}
	P. Schmitteckert,
	Phys. Rev. B \textbf{70}, 121302(R) (2004).

\bibitem{Wang_2003}
	H. Wang, M. Thoss,
	J. Chem. Phys. \textbf{119}, 1289 (2003).

\bibitem{Wang_2011}
	H. Wang, I. Pshenichnyuk, R. H\"artle, M. Thoss, 
	J. Chem. Phys. \textbf{135}, 244506 (2011).

\bibitem{Muehlbacher_2008}
	L. M\"uhlbacher, E. Rabani, 
	Phys. Rev. Lett, \textbf{100}, 176403 (2008).

\bibitem{Muehlbacher_2004}
	L. M\"uhlbacher, J. Ankerhold, C. Escher,
	J. Chem. Phys, \textbf{121}, 12696 (2004).

\bibitem{Weiss_2008}
	S. Weiss, J. Eckel, M. Thorwart, R. Egger,
	Phys. Rev. B. \textbf{77}, 195316 (2008).

\bibitem{Segal_2010}
	D. Segal, A. J. Millis, D. R. Reichman,
	Phys. Rev. B. \textbf{82}, 205323 (2010).


\bibitem{Cohen_2011}
	G. Cohen, E. Rabani,
	Phys. Rev. B. \textbf{84}, 075150 (2011).

\bibitem{Wilner_2013}
	E. Y. Wilner, H. Wang, G. Cohen, M. Thoss, E. Rabani,
	Phys. Rev. B. \textbf{88}, 045137 (2013).

\bibitem{Engel_2007}
	G. S. Engel, T. R. Calhoun, E. L. Read, T.-K. Ahn, T. Mancal, Y.-C. Cheng, 
	R. E. Blankenship, G. R. Fleming,
	Nature (London), \textbf{446}, 782 (2007).

\bibitem{Tanimura_06}
	Y. Tanimura,
	J. Phys. Soc. Jpn. \textbf{75}, 082001 (2006).

\bibitem{Tanimura_09}
	A. Ishizaki, G. R. Fleming,
	J. Chem. Phys. \textbf{130}, 234111 (2009).

\bibitem{Zhang}
	M. Zhang, B. J. Ka, E. Geva,
	J. Chem. Phys \textbf{125}, 044106 (2006).


\bibitem{Stockburger_02}
	J. T. Stockburger, H. Grabert,
	Phys. Rev. B, \textbf{88}, 170407 (2002).

\bibitem{Leggett1987}
	A. J. Leggett, S. Chakravarty, A. T. Dorsey, M. P. A. Fisher, A. Garg, W. Zwerger,
	Rev. Mod. Phys. \textbf{59}, 1 (1987).

\bibitem{Mukamel_1995}
	S. Mukamel,
	\textit{Principles of nonlinear optical spectroscopy}, (Oxford University Press, Oxford, 1995).	

\end{thebibliography}
\end{document}